\documentclass[11pt]{article}
\usepackage{moriond,epsfig}

\usepackage{amssymb,colordvi,gnuplotcolors}

\def\be{\begin{equation}}
\def\ee{\end{equation}}
\def\bea{\begin{eqnarray}}
\def\eea{\end{eqnarray}}

\newcommand{\TOT}{{\mathrm{tot}}}
\newcommand{\HIGGS}{\mathrm{Higgs}}
\newcommand{\BR}{\mathrm{BR}}
\newcommand{\MOD}{{\mathrm{model}}}
\newcommand{\REF}{{\mathrm{ref}}}
\newcommand{\SM}{\mathrm{SM}}
\newcommand{\OBS}{\mathrm{obs}}
\newcommand{\EXPEC}{\mathrm{expec}}

\newcommand{\tb}{\tan \beta}
\newcommand{\gev}{\,\, \mathrm{GeV}}

\newcommand{\sizekey}{\footnotesize}
\newcommand{\size}{\footnotesize}

\begin{document}
\vspace*{4cm}
\title{
{\tt {HIGGSBOUNDS}~\footnote{Online version and code download available at:
{\tt http://www.ippp.dur.ac.uk/HiggsBounds}.}}: %\\
CONFRONTING ARBITRARY HIGGS SECTORS WITH EXCLUSION BOUNDS
FROM LEP AND TEVATRON
\footnote{
Talk presented by K.\ E.\ Williams at ``Rencontres de Moriond -- QCD and High Energy Interactions 2009''
}}

\author{P.\ BECHTLE$^1$, O.\ BREIN$^2$, S.\ HEINEMEYER$^3$, G.\ WEIGLEIN$^4$ AND K.\ E.\ WILLIAMS$^5$}

\address{
$^1$DESY, Notkestrasse 85, 22607 Hamburg, Germany
\\
$^2$Physikalisches Institut,
Albert-Ludwigs-Universit\"{a}t Freiburg,
D-79106 Freiburg, Germany 
\\
$^3$Instituto de Fisica de Cantabria (CSIC-UC), 
Santander, Spain
\\
$^4$Institute for Particle Physics Phenomenology,
Durham University, 
Durham, DH1 3LE, UK
\\
$^5$Physikalisches Institut der Universit\"{a}t Bonn, 53115 Bonn, Germany
}

\maketitle\abstracts{
{\tt HiggsBounds} is a computer code which tests the Higgs sectors of new models against 
the current exclusion bounds from the Higgs searches at LEP and the Tevatron. 
As input, it requires a selection of model predictions, such as Higgs masses, branching ratios, 
effective couplings and total decay widths. {\tt HiggsBounds} then uses the expected and observed 
topological cross section limits from the Higgs searches to determine which points in the 
parameter space have already been excluded at 95\% CL. {\tt HiggsBounds} will be updated 
to include new results as they become available. 
}

\section{Introduction}

The search for Higgs bosons is a major cornerstone
of the physics programmes of past, present and future 
high energy colliders. The LEP and Tevatron 
experiments, in particular, have been able to turn the non-observation of Higgs bosons into constraints on the Higgs
sector, which can be very useful in reducing the available parameter space of particle physics models. Such constraints will continue
to be important far into the LHC era as they will need to be taken into
account in the interpretation of any new physics.

These analyses usually take one of two forms. Dedicated analyses
have been carried out in order to constrain some of the most popular models, such as the SM~\cite{LEP-SM-Higgs-analysis}
and various benchmark scenarios in the MSSM~\cite{LEP-MSSM-Higgs-analysis}. In addition, model-independent 
 limits on the cross sections of individual signal topologies (such as $e^+e^-\to h_iZ\to b\bar{b}Z$) have been published. The former type of analyses include detailed knowledge of the overlap between the
individual experimental searches, and therefore have a high sensitivity, whereas the latter can 
be used to test a wide class of models.

There are certain issues involved with the application of these experimental constraints. The data
is distributed over many different publications and the limits are given
with a variety of normalisations. In the case of the Tevatron, the 
results are also frequently updated. Furthermore, care must be taken when
using more than one experimental analysis to ensure that the resulting exclusion bound has 
the same confidence level (CL) as each individual analysis.

The fortran code {\tt HiggsBounds}~\cite{HBmanual} has been designed to facilitate the task of comparing Higgs sector 
predictions with existing exclusion limits, thus allowing the user to quickly and conveniently 
check a wide variety of models against the state-of-the-art results from Higgs
searches. 

\section{Outline of the program}

%\begin{enumerate}
%\item 
The user provides the Higgs sector predictions of the model under
consideration. For each neutral Higgs boson $h_i \; (i=1,\ldots,n_\HIGGS)$ in the model,
this will usually include the mass, total decay width, branching ratios and Higgs production cross sections:
\bea
\label{basic input}
& M_{h_i} \,, 
\Gamma_{\TOT}(h_i)\,,  
\BR_\MOD(h_i\to ...)\,, 
 \frac{\sigma_\MOD(P)}{\sigma_{\REF}(P)}&\, .
\eea
Where it exists, $\sigma^\SM(P)$ is used as the reference cross section. Variations on this input format are offered, as described in detail in the {\tt HiggsBounds} manual~\cite{HBmanual}. The {\tt HiggsBounds} package includes 
sample programs which demonstrate how {\tt HiggsBounds}
can be used in conjunction with the 
widely used MSSM Higgs sector programs 
{\tt FeynHiggs}~\cite{feynhiggs-long} and 
{\tt CPsuperH}~\cite{cpsh}.

\begin{table}[t]
\caption{
\label{LEP-search-topologies}
LEP and Tevatron analyses used by 
{\tt HiggsBounds}. $l$ or $l'$ indicates an electron or a muon, and $\dag$ indicates analyses which combine processes using SM assumptions. In this notation,  $m_{h_k} > m_{h_i}$.
}
\vspace{0.4cm}
\begin{center}
\begin{tabular}{|ll|}
\hline
{\size $e^+e^-\to (h_k)Z\to (b \bar{b})Z$ }
&{\size $p\bar p \to Z H \to l^+l^- b\bar b$ ( CDF,D\O\ ) }\\
{\size $e^+e^-\to (h_k)Z\to (\tau^+ \tau^-)Z$ }
&
{\size $p\bar p \to W H \to l\nu b\bar b$ ( CDF,D\O\ )}\\
{\size $e^+e^-\to (h_k\to h_i h_i)Z\to (b \bar{b} b \bar{b})Z$ }
&
{\size $p\bar p \to W H \to l\nu b\bar b$ ( CDF,D\O\ )}\\
{\size $e^+e^-\to (h_k\to h_i h_i)Z\to (\tau^+ \tau^- \tau^+ \tau^-)Z$ }
&{\size $p\bar p \to H \to W^+W^- \to l^+ l'^-$ ( CDF,D\O\ )}\\
{\size $e^+e^-\to (h_k h_i)\to (b \bar{b} b \bar{b})$ }
&
{\size $p\bar p \to H \to\gamma\gamma$ ( CDF,D\O\ )}
\\
{\size $e^+e^-\to (h_k h_i)\to (\tau^+ \tau^- \tau^+ \tau^-)$ } 
&
{\size $p\bar p \to H \to \tau^+\tau^-$ ( CDF,D\O\ )}
\\
{\size $e^+e^-\to (h_k\to h_i h_i)h_i\to b \bar{b} b \bar{b}b \bar{b}$} 
&
{\size $p\bar p \to b H, H \to b\bar b$ ( CDF,D\O\ )}
\\
{\size $e^+e^-\to (h_k\to h_i h_i)h_i\to \tau^+ \tau^- \tau^+ \tau^-\tau^+ \tau^-$ }
&
{\size $p\bar p \to W H/Z H \to b\bar b + E_T^{\mathrm{miss.}}$ ( CDF,D\O\ )$\dag$}
\\
{\size $e^+e^-\to (h_k\to h_i h_i)Z\to (b \bar{b}\tau^+ \tau^-)Z$ }
&
{\size $p\bar p \to H/HW/HZ/H \mathrm{\ via VBF}$,}
\\
{\size $e^+e^-\to (h_k\to b \bar{b})(h_i\to \tau^+ \tau^-)$ }
&
{\size  \,\,\,\,\,\, \,\,\,  \, $H\to \tau^+\tau^-$ 
                                                      ( CDF )$\dag$}
\\
{\size $e^+e^-\to (h_k\to \tau^+ \tau^-)(h_i\to b \bar{b})$ }
&
{\size combined Tevatron analyses for the SM Higgs$\dag$}
\\
\hline
\end{tabular}
\end{center}
\end{table}

%\item 
A list of the experimental analyses~\cite{LEP-MSSM-Higgs-analysis,expt-notes,D0-0805-2491} currently included in {\tt HiggsBounds}
is given in Table~\ref{LEP-search-topologies}. These include results from both LEP and the Tevatron and consist of tables of expected (based on Monte Carlo simulations with no signal) and observed 95\% CL cross section limits, with a variety of normalisations. The list mainly consists of analyses for which model-independent limits were published.
However, we also include some dedicated analyses carried out for the case of the SM. These analyses are only considered
if the Higgs boson in question would appear sufficiently `SM-like' to this analysis. Roughly speaking, this requires that
the ratios of all involved couplings to the SM couplings are approximately equal~\cite{HBmanual}. 

For each Higgs process $X$ (here, we treat each combination of Higgs bosons in each experimental analysis as a separate $X$), {\tt HiggsBounds} uses the input to calculate the quantity $Q_\MOD(X)$, which, up to a normalisation factor, is the predicted cross section for $X$. 

%\footnote{If two Higgs bosons have a narrow separation $\delta M_h=M_{h_i}-M_{h_j}$, then their predicted cross sections are added. The value $\delta M_h$ can be varied by the user separately for LEP and Tevatron analyses.}

The normalisation is carried out using SM predictions for 
Higgs boson production cross sections
and decay branching ratios from
{\tt HDECAY}~\cite{hdecay} 3.303, 
the TEV4LHC Higgs Working Group~\cite{TEV4LHCWG-Higgs-CS-plus,HJET}, 
{\tt VFB@NLO}~\cite{VBFatNLO},
{\tt HJET}~\cite{HJET} 1.1 and
dedicated calculations of our own~\cite{HBmanual}. 
%\item 

In order to ensure the correct statistical interpretation of the results,
it is crucial to only consider the experimentally observed limit for one particular $X$. Therefore, {\tt HiggsBounds} must first determine $X_0$, which is defined as the process $X$ with the highest
statistical sensitivity for the model point under consideration. In order to do this, the program uses the tables of expected experimental limits to obtain a quantity $Q_\EXPEC$ corresponding to each $X$. The process with the largest value of $Q_\MOD/Q_\EXPEC$ is chosen as $X_0$.

{\tt HiggsBounds} then derives a value for $Q_\OBS$ for this process $X_0$, using the appropriate table of experimentally observed limits. If
\begin{equation}
\frac{Q_\MOD(X_0)}{Q_\OBS(X_0)} > 1
\,,
\label{eq:modvsobs}
\end{equation} 
{\tt HiggsBounds} concludes that this particular parameter point is excluded at 95 \% CL.
%\end{enumerate}
%

In order to use {\tt HiggsBounds}, the narrow-width approximation 
must be valid for each Higgs boson described in the input. This is because
the experimental exclusions bounds currently utilised within {\tt HiggsBounds} have
all been obtained under this approximation.
We intend to include width-dependent limits into {\tt HiggsBounds} in the future,
where they are provided by experimental collaborations in a model-independent format 
(such as those provided in Ref.\cite{D0-0805-2491}).

\begin{figure}[tb]
\begin{center}
\caption{Coverage of the LEP Higgs searches in the $M_{H_1}$--$\tb$ plane
of the CPX scenario, where $M_{H_1}$ is the lightest neutral Higgs boson and $\tb$ is the ratio of vacuum expectation values. Left: the LEP processes 
predicted to have the highest statistical sensitivity at each parameter point.
Right: the parameter regions excluded
at the 95\% CL (green (dark grey) = excluded, 
white = unexcluded, light grey = theoretically inaccessible)
\label{fig:CPX}
}
\begin{tabular}{ccc}
\begin{minipage}{.33\linewidth}
{\sizekey
Key to processes (left-hand graph):
\bea
\mathrm{red}(\GNUPlotA{\blacksquare})=(h_1 Z)&\to&(b \bar b Z)\nonumber
\nonumber\\
\mathrm{blue}(\GNUPlotC{\blacksquare})=(h_2  Z)&\to&(b \bar b Z)
\nonumber\\
\mathrm{white}(\square)=(h_2  Z)&\to&(h_1 h_1 Z)
\nonumber\\
&\to&(b \bar b b \bar b Z)
\nonumber\\
\mathrm{cyan}(\GNUPlotE{\blacksquare})=(h_2 h_1)&\to&(b \bar b b \bar b)
\nonumber\\
\mathrm{yellow}(\GNUPlotF{\blacksquare})=(h_2 h_1)&\to&(h_1 h_1 h_1)
\nonumber\\
&\to&(b \bar b b \bar b b \bar b)
\nonumber\\
\vspace{-0.1cm}
\mathrm{purple}(\Purple{\blacksquare})=\mathrm{other}&&
\nonumber
\eea
}
\end{minipage}
&\raisebox{-.15\linewidth}{
\hspace{-0.4cm}\includegraphics[width=.33\linewidth,angle=0]{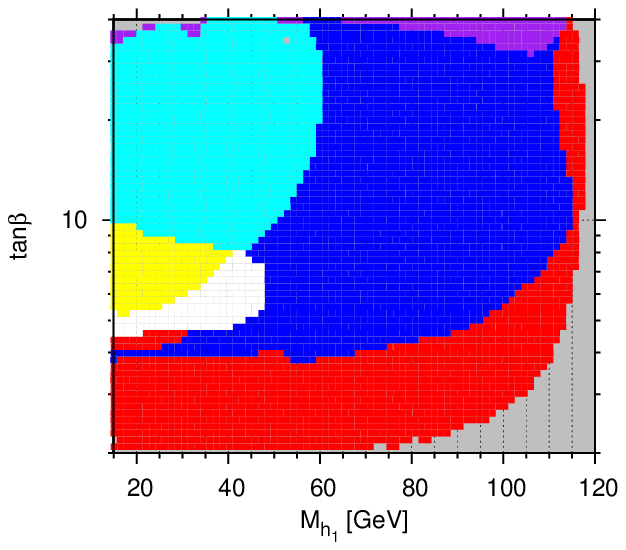}
}&\raisebox{-.15\linewidth}{
\hspace{-0.6cm}\includegraphics[width=.33\linewidth,angle=0]{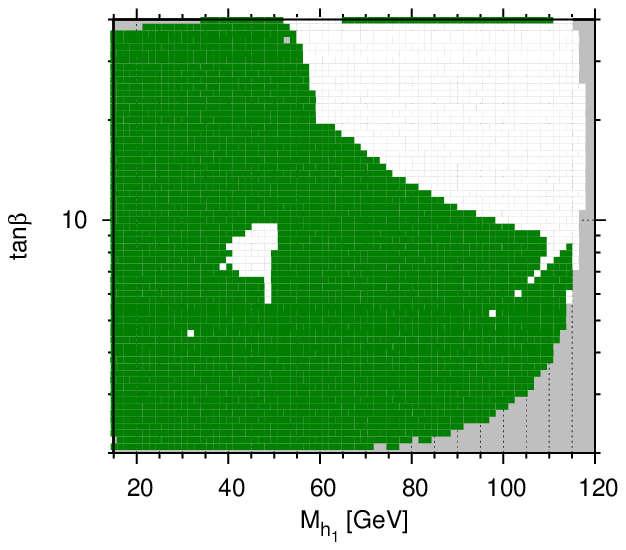}
}\\
%hello
%
%
%
\end{tabular}
\end{center}
\vspace{-0.4cm}
\end{figure}

\section{Numerical example: the CPX scenario}

We will illustrate some of the main features of {\tt HiggsBounds} using an example from Ref.~\cite{Williams:2007dc} (with the modifications described in Ref.~\cite{HBmanual}). The CPX scenario was one of the MSSM scenarios which were investigated in detail by the LEP Higgs Working Group~\cite{LEP-MSSM-Higgs-analysis}. It is phenomenologically interesting because it introduces large CP-violating phases, which induce mixing in the neutral Higgs sector, resulting in weaker exclusions than those obtained for the real MSSM. However, since this original analysis, there have been relevant theoretical advances~\cite{Heinemeyer:2007aq,Williams:2007dc}, which can have a large numerical effect on the Higgs sector of the CP-violating MSSM. Therefore, {\tt HiggsBounds} was employed to investigate the effect of these new results on the amount of CPX parameter space which can be excluded by current Higgs search data.

From Fig.~\ref{fig:CPX} (right), it can be seen that, although substantial regions of CPX parameter space can be excluded (green), there are significant regions which remain unexcluded (white), including an unexcluded region (the `CPX hole') at a lightest Higgs mass $M_{h_1}\sim 45 \gev$, qualitatively confirming the result of the original analysis~\cite{LEP-MSSM-Higgs-analysis}. In addition, the use of {\tt HiggsBounds} allows a greater understanding of the theoretical influences on the exclusions. For example, it can be seen from Fig.~\ref{fig:CPX} (right) that the process $e^+e^-\to(h_1 Z)\to(b \bar b Z)$ (red), which is usually the most effective at excluding areas of MSSM parameter space, has the highest statistical sensitivity only in regions with low $\tb$ and/or high $M_{h_1}$. This is because the coupling of the lightest Higgs to two $Z$ bosons is suppressed in the other regions, therefore reducing the $h_1$ Higgsstrahlung production cross section. It is also interesting to note that, near to the CPX hole, the processes with the highest statistical sensitivity all directly involve the decay of the second heaviest neutral Higgs $h_2$. Therefore, it can be inferred that variations in the partial decay widths of the dominant decay modes (in this case, the Higgs cascade decay $h_2\to h_1 h_1$ and the decay to b-quarks $h_2 \to b \bar{b}$) will affect the size and position of the CPX hole, as is indeed the case~\cite{Williams:2007dc}. 

In conclusion, the program {\tt HiggsBounds} provides a convenient way to compare theoretical Higgs sector predictions with the current exclusion bounds from LEP and the Tevatron, in a way that maintains the statistical interpretation of the exclusion limit and gives extra insight into phenomenological influences on the result.

%\begin{figure}
%\rule{5cm}{0.2mm}\hfill\rule{5cm}{0.2mm}
%\vskip 2.5cm
%\rule{5cm}{0.2mm}\hfill\rule{5cm}{0.2mm}
%%\psfig{figure=filename.ps,height=1.5in}
%\caption{Radiative (off-shell, off-page and out-to-lunch) SUSY Higglets.
%\label{fig:radish}}
%\end{figure}

\section*{Acknowledgements}
K.~W.\ would like to thank the organizers of 
``Rencontres de Moriond'' for an enjoyable and inspiring conference.
We are grateful for the valuable assistance of A. Read, P. Igo-Kemenes,
M.~Owen, T.~Junk, M.~Herndon and S. Pagan Griso. This work has been supported in part by the Helmholtz Alliance HA-101, the Marie-Curie Research Training Network under contract MRTN-CT-2006-035505 and MRTN-CT-2006-035657 and by the Helmholtz Grant VH-NG-303.

\end{document}